\documentclass[10pt,twocolumn,letterpaper]{article}

\usepackage{iccv}
\usepackage{times}
\usepackage{epsfig}
\usepackage{graphicx}
\usepackage{amsmath}
\usepackage{amssymb}
\usepackage{subcaption}
\usepackage{booktabs}
\usepackage{multirow}
\usepackage{scrlayer-scrpage}
\usepackage[section,ruled]{algorithm}
\usepackage{algorithmic}
\usepackage[breaklinks=true,bookmarks=false]{hyperref}

\newcommand\Bern{\mathord{\mathrm{Bern}}}

\iccvfinalcopy % *** Uncomment this line for the final submission

\def\iccvPaperID{10702} % *** Enter the ICCV Paper ID here

\ificcvfinal\fi

\begin{document}

%%%%%%%%% TITLE
\title{Deep Hashing with Hash Center Update for Efficient Image Retrieval}

\author{Abin Jose, Daniel Filbert, Christian Rohlfing, and Jens-Rainer Ohm\\
Institut für Nachrichtentechnik\\
RWTH Aachen, Germany\\
{\tt\small jose@ient.rwth-aachen.de}}

\maketitle
\ificcvfinal\thispagestyle{empty}\fi

%%%%%%%%% ABSTRACT
\begin{abstract}
In this paper, we propose an approach for learning binary hash codes for image retrieval. Canonical Correlation Analysis (CCA) is used to design two loss functions for training a neural network such that the correlation between the two views to CCA is maximized. The first loss, maximizes the correlation between the hash centers and learned hash codes. The second loss maximizes the correlation between the class labels and classification scores. A novel weighted mean and thresholding based hash center update scheme is proposed for adapting the hash centers in each epoch. The training loss reaches the theoretical lower bound of the proposed loss functions, showing that the correlation coefficients are maximized during training and substantiating the formation of an efficient feature space for image retrieval. The measured mean average precision shows that the proposed approach outperforms other state-of-the-art approaches in both single-labeled and multi-labeled image datasets.
\end{abstract}

%%%%%%%%% BODY TEXT
\section{Introduction}
Due to their excellent feature extraction capabilities,
deep neural networks have become state-of-the-art in feature extraction and are the basis of most Content Based
Image Retrieval methods [42]. The evaluation of similarity between two images would be performed by computing
a predefined distance measure between the feature vectors
in the feature space. Besides high retrieval quality, both
efficiency and speed are also essential requirements for a
good retrieval system. Due to the dramatic and continuous
growth of image datasets, evaluation of Euclidean distances
for subsequent ranking and nearest neighbor search has become too computationally expensive or even infeasible [35].
In addition, the feature vectors are high-dimensional, which
increases the computational complexity exponentially. A
solution for this problem would be using compact binary
representations. In general, when using neural networks
as a feature extractor for image retrieval, two approaches
can be considered for obtaining binary codes. In a first
approach, an Euclidean feature space can be optimized resulting in feature vectors having well suited properties for
retrieval when binarized [19]. Subsequent quantization and
binarization of the Euclidean feature space is then mainly to
be concerned with minimizing the information loss as well
as preserving similarity information in the obtained binary
codes. In the second approach, a deep neural network can
be trained to directly produce effective binary features [3].
In this case, obtaining binary values also requires the use
of an activation function which maps to a binary space in
the last layers of the network. We have adopted the second approach in this paper. To overcome the challenges of
high computational cost and lack of search speed, Approximate Nearest Neighbour [39] (ANN) search is a common
alternative approach which offers more efficiency and sufficiently high accuracy for many practical applications [33].
A widely used form of ANN search is hashing, where a
mapping is found to create a lower-dimensional representation of the actual data while preserving the similarity between data points in the new domain accurately.
Hashing is in general classified into data-independent
approaches and data-dependent approaches. Data independent approaches [22, 17, 29] mainly rely upon random projections to generate the hash functions. Locality-sensitive
hashing (LSH), is [6] a popular data-independent approach,
which was used in [15] and [9] to solve the ANN problem
while avoiding the curse of dimensionality inherently associated with exact nearest neighbor searches for data in their
original metric spaces. Recent methods mainly concentrate
on, data-dependent approaches which are mainly categorized as unsupervised and supervised methods. We refer
readers to [34] for an extensive survey. For learning hash
functions, unsupervised approaches [21, 8, 7, 26, 14, 13]
use various metrics to supervise the learning. In contrast,
the supervised approaches utilize the semantic labels of the
training data. In recent years, deep supervised hashing has
achieved good retrieval performance [36, 3, 24, 18, 2, 25,
37, 4, 2, 24, 18]. Especially, pair-wise and triplet hashing
approaches [3, 41, 36, 25, 37] have shown promising results. However, these approaches require a lot of time to
sample enough pairs or triplets. The loss function depends on the distance calculation between similar and dissimilar
pairs and does not utilize the entire feature statistics during
training.
\subsection{Related work}
Quite recently, Canonical Correlation Analysis (CCA) was used such that the correlation between the feature vectors and label vectors was maximized \cite{jose2020optimized}. The learned feature space was then binarized using the popular ITQ \cite{gong2012iterative} approach. Here, statistical properties of the feature space is taken into consideration during the training.\par Yuan et al. proposed a new global similarity metric
which they called central similarity in \cite{yuan2020central}. By applying this new metric, hash values of similar data points are encouraged to approach a
common center, where as pairs of dissimilar hash codes converge to distinct
centers in the Hamming space. There are two systematic approaches proposed in \cite{yuan2020central} to
generate hash centers fulfilling the above condition:
One leverages the characteristics of the Hadamard matrix, thus obtaining
hash centers with maximal mutual Hamming distance, and the other uses random
sampling from a Bernoulli distribution when the bit length is not a power of $2$. Generating hash centers from the rows of the Hadamard matrix has several appealing properties: Firstly, it is a binary matrix
with elements of value $\{+1, -1\}$ which makes the generation of the hash centers in the
Hamming space straightforward. Furthermore, it is a square
matrix of size $K \times K$ with $K = 2^n, n \in \mathbb{N}$ being a power of $2$
which leads to hash centers with a common amount of bits in the hash codes. In addition, its row vectors are mutually orthogonal.
Learning the hash function requires training data to be associated with the
generated hash centers in order to reflect semantic information in the
Hamming space. There are as many hash centers generated as there are semantic labels in the dataset. However, since each sample can contain one or more categories for multi-labeled data, a majority voting is proposed in order to account for the transitive similarity of data points sharing multiple labels. For training, a central similarity loss is defined which 
utilizes the binary cross entropy loss to measure the similarity between hash
code outputs of the network and its corresponding semantic hash centers. To avoid optimization difficulties implied by the binary valued hash centers, a quantization loss was introduced based on a bi-modal Laplacian prior, as in
\cite{zhu2016deep}, with additional smoothing. However, a major problem in this approach is that the hash centers are not updated even though the feature space changes during training.\par Another interesting work in a similar direction was proposed by Hong et al. in \cite{hong2020image}.
Here, Linear Discriminant Analysis characteristics are trained directly on the hash codes, thus
enforcing the deep network to produce hashes which have a small intra-class
variance while also having a high inter-class variance.
The proposed method updates the hash centers
during deep hashing training. Here, the
natural problem arises that hash centers are desired to be binary, such
that the CNN features are encouraged to be discrete and are desired to be real valued at the same time, such that gradient descent optimization is feasible
~\cite{hong2020image}. Therefore, a distinct treatment of those hash centers is designed
depending on whether performing a forward pass or backward pass step during training is implemented.
\section{This paper}
Inspired by the idea of updating hash centers during training ~\cite{hong2020image}, an alternative
method of hash center update based on the weighted mean of the hash values is proposed in our paper, which aims to reflect the movement
of the formed clusters in the Hamming space during training. The network was trained using a CCA-based loss formulation such that the correlation between the hash codes and hash centers is maximized along with the correlation of classification scores and class labels.
The major contributions of this paper are summarized below:
\begin{itemize}
	\item First, initial hash centers, around which the hash codes are clustered, are selected as proposed in \cite{yuan2020central}. A novel weighted mean and thresholding based hash center update scheme is proposed for both single-labeled and multi-labeled images.
	\item The loss function is formulated using CCA such that the generated hash codes and hash centers have maximum correlation. This loss function is combined with the CCA-based classification loss as proposed in \cite{jose2020optimized} which maximizes the correlation of classification scores with the class labels.
	\item The theoretical lower bound is determined for both loss functions based on the rank of the two views of CCA, and an optimum regularization factor is chosen to combine the two loss functions.
\end{itemize}\par
Experiments were conducted on single-labeled dataset CIFAR-10 \cite{krizhevsky2009learning}, as well as
multi-labeled datasets MS-COCO \cite{lin2014microsoft} and NUS-WIDE
\cite{chua2009nus} and the retrieval performance of the hash codes is evaluated for different bit lengths. The training and test curves, precision-recall curves, and t-SNE \cite{t-sne} plots of the generated feature space were plotted and discussed. Mean average precision was computed and the performance is compared with other supervised hashing approaches such as DPSH~\cite{li2015feature}, DCCH~\cite{jose2020optimized}, CSQ~\cite{yuan2020central}, DSDH~\cite{li2017deep}, DDSH~\cite{jiang2018deep}, DTSH~\cite{wang2016deep}, LDH~\cite{hong2020image}, HashNet~\cite{cao2017hashnet}, DHN~\cite{zhu2016deep}, DNNH~\cite{lai2015simultaneous}, and CNNH~\cite{xia2014supervised}.\par 
The paper is organized as follows: Section \ref{sec:dcsh-motivation} explains the network architecture, and loss function. The hash center update is discussed in Section \ref{sec:dcsh-hash_center_update}. Experimental results are discussed in Section \ref{sec:expts} and concluding remarks are drawn in Section \ref{sec:conc}.
\section{Proposed approach} %: CCA Characteristics Directly In Binary Space}
\label{sec:dcsh-motivation}
An efficient feature space for image retrieval reflects the semantic information
contained in its represented images. Deep hashing is used in the proposed Deep Central Similarity Hashing (DCSH) method, which directly learns compact binary hash codes having high correlation with the hash centers representing the image categories. The hash centers are updated during training such that they adapt to the changes in the feature space.
\subsection{Network architecture}
\label{sec:dcsh-network_architecture}
\begin{figure*}[!htb]
	\center
	\includegraphics[width=1\textwidth]{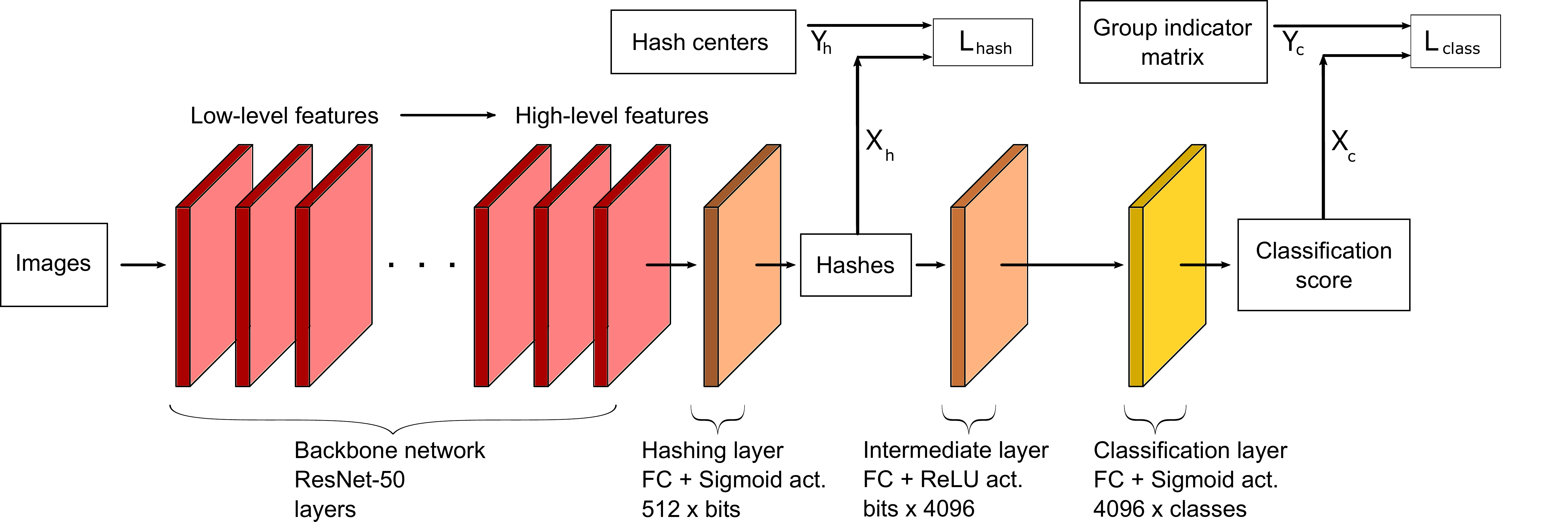}
	\caption{ {Overview of the proposed Deep Central Similarity Hashing network architecture.
			ResNet layers according to \cite{he2016deep} are used as the backbone network
			for basic feature extraction. Both hashing and classification layer,
			consist of a fully connected (FC) layer with subsequent sigmoid activation. The
			intermediate layer comprises a fully connected layer with subsequent ReLU
			activation. Bits indicate the bit length of the hash code. Classes indicates the number of categories in the dataset.}}
	\label{fig:dcsh-network_architecture}
\end{figure*}
The network architecture of the proposed approach is given in Fig. \ref{fig:dcsh-network_architecture}. The
Residual network \cite{he2016deep} is used as the basic feature extractor which was pretrained on the ImageNet \cite{Imagenet} dataset. Followed by the residual layers, a hashing layer
consisting of a fully connected layer and subsequent sigmoid activation
function is used to generate
hashes corresponding to the given input images. The output dimension of this layer is
therefore equal to the required number of bits. An intermediate layer is used subsequently to the hashing layer to
generate high output dimensionality from the input hash codes. This intermediate layer is required, since the loss function used in this architecture, $L_{\mathrm{DCCF}}$ (see Section \ref{sec:dcsh-loss_formulation}), is a dimensionality reduction method. A higher number of input dimensions exceeding the number of distinct classes in the
dataset is required as input to this layer. There are two losses in the proposed approach:
1) $L_{\mathrm{hash}}$ correlating the hashing outputs with the semantic hash centers. 2) $L_{\mathrm{class}}$ correlating the classification scores with the label information of the dataset.
The final training loss, $L_{\mathrm{DCSH}}$ then takes both losses into account for the optimization. 
%\begin{table}[htb!]
%	\begin{center}
%			\caption{Deep Central Similarity Hashing network architecture}
%		\begin{tabular}{  l | l  }
%			\toprule
%			\textbf{Block}            & \textbf{Details}                                          \\
%			\midrule
%			\multirow{2}{*}{ResNet50} & Convolutional Layers                                      \\
%			& as outlined in Table \ref{tab:dcsh-resnet50_architecture} \\
%			\midrule
%			& Fully Connected                                           \\
%			Hashing Layer             & Sigmoid Activation                                        \\
%			& $(512 \times \#bits)$                                     \\
%			\midrule
%			& Fully Connected                                           \\
%			Intermediate Layer        & ReLU Activation                                           \\
%			& $(\#bits \times 4096)$                                    \\
%			\midrule
%			& Fully Connected                                           \\
%			Classification Layer      & Sigmoid Activation                                        \\
%			& $(4096 \times \#classes)$                                 \\
%			\bottomrule
%		\end{tabular}
%		\label{tab:dcsh-network_architecture}
%	\end{center}
%\end{table}
\subsection{Hash code generation}
\label{subsec:dcsh-binary_codes_generation}
In proposed DCSH, hashing values are generated by the output of the
hashing layer. It contains a sigmoid activation function which produces hashes $\vec{x}_h
\in [0, 1]^B$ with $B$ elements, and $B$ being the number of required bits for each of
the $N$ images in the dataset.  After successful training of the
network, the values of the hashing outputs are likely to be either close to
$0$ or $1$. An additional thresholding $\tau(\cdot)$ at $0.5$ is performed
on each element to obtain the desired binary code vectors.

\subsection{Loss formulation}
\label{sec:dcsh-loss_formulation}
The proposed approach uses Canonical Correlation Analysis (CCA) at two output layers of the network to formulate the training loss as shown in Fig. \ref{fig:dcsh-network_architecture}. CCA
aims to find transformations of two input views which maximally correlates
their mapped representations. DCCH \cite{jose2020optimized} uses CCA such that a
neural network can be trained to generate non-linear mappings of its input
which maximally correlate to a given target. In this paper, this loss formulation is used to optimize the outputs of a neural
network from two layers simultaneously. The layers from which the loss is calculated is shown in Fig. \ref{fig:dcsh-network_architecture}.
%\begin{figure}[!htb]
%  \begin{center}
%    \input{figures/deep_central_similarity_hashing_loss_block_diagram}
%    \caption{Loss function design in Deep Central Similarity Hashing}
%    \label{fig:dcsh-loss_block_diagram}
%  \end{center}
%\end{figure}
For two data views $\mathbf{X}$ and $\mathbf{Y}$, CCA optimizes projections
$\vec{a}$ and $\vec{b}$ which maximize the correlation $\rho$ between the
projected inputs as:
\begin{equation}
	\begin{aligned}
		\rho(\vec{a}^*, \vec{b}^*) &= \max_{\vec{a}, \vec{b}} \mathrm{corr}(\vec{a}^T \mathbf{X}, \vec{b}^T \mathbf{Y}) \\
		                           & = \max_{\vec{a}, \vec{b}}
		\frac{\vec{a}^T \mathbf{\Sigma_{XY}} \vec{b}}
		{\sqrt{\vec{a}^T \mathbf{\Sigma_{XX}} \vec{a} \quad \vec{b}^T \mathbf{\Sigma_{YY}} \vec{b}}} \\
		&\text{s.t. } \vec{a}^T \mathbf{\Sigma_{XX}} \vec{a} = \vec{b}^T \mathbf{\Sigma_{YY}} \vec{b} = 1 \text{ ,}
	\end{aligned}
\end{equation}
where, $\mathbf{\Sigma_{XX}}, \mathbf{\Sigma_{XY}},$ and $\mathbf{\Sigma_{YY}}$ denotes the covariances,
$\mathrm{cov}{\mathbf{(X,X)}}, \mathrm{cov}{\mathbf{(X,Y)}}, $ and $ \mathrm{cov}{\mathbf{(Y,Y)}}$ respectively.
This optimization problem can be solved by using a Singular Value
Decomposition, shown by Mardia et al. in
\cite{mardia1979multivariate}. DCCF followed this approach to formulate the loss
function as:
\begin{equation}
	\label{eq:dcsh-dccf_loss}
	\begin{aligned}
		L_{\mathrm{DCCF}} &= - \sum_{i=1}^{k} \sigma_i = - \sum_{i=1}^{k} \rho_i \\
	\end{aligned}
\end{equation}
with $k$ largest singular values $\sigma_i$ of the matrix $\mathbf{K}$ defined by
\begin{equation}
	\label{eq:dcsh-mat_K_def}
	\mathbf{K} := \mathbf{\Sigma}_{\mathbf{XX}}^{-1/2} \mathbf{\Sigma}_{\mathbf{XY}} \mathbf{\Sigma}_{\mathbf{YY}}^{-1/2} \text{ .}
\end{equation}
The equivalence between the correlation coefficients and the singular values is also proven in \cite{mardia1979multivariate}. 
Accordingly, the training loss minimizes the negative sum of correlation
coefficients. Since the maximum positive correlation can reach a value of $1$, the lower bound of the DCCF loss $L_{\mathrm{DCCF}}$ will be
$-k$. Furthermore, $k$ can only be as high as the rank of matrix $\mathbf{K}$. For
input vectors $\mathbf{X}$ and target data view $\mathbf{Y}$, the upper bound of $k$
is determined by:
\begin{equation}
	\label{eq:dcsh-dccf_loss_kmax}
	k_{\mathrm{max}} = \text{rank}(\mathbf{K}) = \min(\text{rank}(\mathbf{X}) \text{, rank}(\mathbf{Y})) - 1 \text{ .}
\end{equation}
The subtraction of 1 from either $\text{rank}(\mathbf{X})$ or
$\text{rank}(\mathbf{Y})$ is due to the inherent subtraction of the mean for the
covariance matrices in $\mathbf{K}$. The loss function of DCSH consists of two CCA evaluations. Both components, the hashing loss and the
classification loss, are discussed next.\par
\textbf{Hashing loss:} First, the DCCF loss is evaluated by using the outputs of the hashing layer
$\mathbf{X}_\mathrm{h}$ with their corresponding semantic hash centers $\mathbf{Y}_\mathrm{h}$. These
hash centers act as targets in Hamming space towards which the respective
output hashes of the network should converge.  The size of the two data views
$\mathbf{X}_\mathrm{h} \in [0, 1]^{M \times B}$ and $\mathbf{Y}_\mathrm{h} \in \{0, 1\}^{M \times B}$ is determined by the batch
size $M$ and the number of bits $B$ in the binary codes.  Therefore, the
hashing loss can be formulated as:
\begin{equation}
	L_{\mathrm{hash}} = L_{\mathrm{DCCF}}(\mathbf{X}_\mathrm{h}, \mathbf{Y}_\mathrm{h}) \text{ .}
\end{equation}
Since each hash center used in the target view $\mathbf{Y}_\mathrm{h}$ represents one of
the $C$ categories in the underlying dataset, it consists of
only at most $C$ distinct rows. Assuming that $M > C$, $\text{rank}(\mathbf{Y}_\mathrm{h}) = \min (B, C)$.  Therefore, the maximal
number of correlation coefficients according to Eq.
\eqref{eq:dcsh-dccf_loss_kmax} is:
\begin{equation}
	\begin{aligned}
		k_{\mathrm{max}} &= \min(\min(B, M), \min(B, C)) - 1 \\		
		&= \min(B, C) - 1 \text{ , for $M > B$ and $M > C$.}
	\end{aligned}
\label{old_equ}
\end{equation}
According to Eq. \eqref{eq:dcsh-dccf_loss}, this effectively means that during
training the negative sum of $\min(B, C) - 1$ correlation coefficients is
minimized. As the correlation maximally can reach a value of $1$, the lower
bound of the hashing loss goes to $ - (\min(B, C) - 1)$.\par
\textbf{Classification loss:}
The second component of the proposed DCSH loss function is a
classification loss, which performs a CCA of the classification scores
$\mathbf{X}_\mathrm{c}$ from the output of the final network layer and a target group
indicator matrix $\mathbf{Y}_\mathrm{c}$ (see Fig. \ref{fig:dcsh-network_architecture}).
In this context the first data view $\mathbf{X}_\mathrm{c} \in \mathbb{R}^{M\times L}$ is defined as a matrix consisting of $M$ rows each denoting an
$L$-dimensional feature representation $\vec{x}^T \in \mathbb{R}^L$ of a sample
of the current batch with $M$ images. As a second data view, a so called group
indicator matrix $\mathbf{Y}_\mathrm{c} \in \{0, 1\}^{M \times C}$ is used with $M$ rows of
label vectors $\vec{y} = [y_1, ..., y_C]$ with $C$ entries (with a value $1$ indicating the class associated to the respective image and a value $0$ otherwise).
This definition can directly be extended for multi-class labels, such that each row
of the group indicator matrix $\mathbf{Y}_\mathrm{c}$ indicates the associated classes in the
corresponding data sample \cite{jose2020optimized}.
Note that the classification layer uses a sigmoid activation instead of the
usual softmax. This is because the approach aims to be directly extensible to
multi-label datasets using the same CCA evaluation. As group indicator matrices
are used as the second view, sigmoid activation is an appropriate choice
encouraging the network to produce outputs close to $1$ for classes
contained in the image and $0$ otherwise. Here, the size of the
two used data views $\mathbf{X}_\mathrm{c} \in [0, 1]^{M \times C}$ and $\mathbf{Y}_\mathrm{c} \in \{0,
1\}^{M \times C}$ is determined by the batch size $M$ and the number of
distinct classes $C$ in the underlying dataset. Given these data views, the
classification loss is formulated as:
\begin{equation}
	L_{\mathrm{class}} = L_{\mathrm{DCCF}}(\mathbf{X}_\mathrm{c}, \mathbf{Y}_\mathrm{c}) \text{ .}
\end{equation}
Assuming a higher batch size than distinct classes $M > C$, the maximal
number of correlation coefficients according to Eq.
\eqref{eq:dcsh-dccf_loss_kmax} results:
\begin{equation}
	\begin{aligned}
		k_{\mathrm{max}} 	&= \min(M, C) - 1 \\
		&= C - 1 \text{ , for $M > C$.}
	\end{aligned}
\label{new_equ}
\end{equation}
The
lower bound of the classification loss then comes to a value of $- (C - 1)$.\par
\textbf{Loss combination and normalization:}
\label{subsec:dcsh-loss_combination}
The final training objective of the proposed deep hashing method is designed by
a linear combination of the two losses introduced above as:
\begin{equation}
 L_{\mathrm{DCSH}} = L_{\mathrm{hash}} + \alpha  L_{\mathrm{class}} \text{ .}
 \end{equation}
The regularization factor $\alpha$ balances the contribution of both components
for the final optimization objective. By using the theoretical lower bounds of
hashing and classification loss as shown in Eq. \eqref{old_equ} and Eq. \eqref{new_equ}, an equal weighting of their
contributions can be computed as:
\begin{equation}
	\label{eq:dcsh-alpha_normalizing}
	\alpha = \frac{\min(B, C) - 1}{C - 1} \text{ .}
\end{equation}
Setting $\alpha$ to this value results in the theoretical lower bounds of the
hashing and classification loss both becoming equal. However, it
has been beneficial in subsequent experiments to choose $\alpha$ in each case of $B$ and $C$ as:
\begin{equation}
 \alpha = \frac{B - 1}{C - 1} \text{ .}
 \end{equation}
Therefore, in case of datasets for which $C < B$, the
classification loss is emphasized. This occurs for all single-labeled datasets. Experiments on
multi-labeled datasets all satisfy $C > B$ and thus, $\alpha$
normalizes the contribution of classification and hashing loss in the final
training objective. The theoretical lower bound of the
combined DCSH loss can then be determined as follows:
\begin{equation}
	\label{eq:dcsh-lower_bound_combined}
	\begin{aligned}
		\min L_{\mathrm{DCSH}} &= \min L_{\mathrm{hash}} + \alpha \min L_{\mathrm{class}} \\
		%&= - (\min(B, C) - 1) - \frac{B - 1}{C - 1} (C - 1) \\
		&= - (\min(B, C) - 1) - (B - 1) \text{ .}
	\end{aligned}
\end{equation}
\section{Hash center update during training}
\label{sec:dcsh-hash_center_update}
An important part of the training procedure in DCSH is
to provide good semantic hash centers, which are updated in each epoch. According to \cite{yuan2020central}, either the Hadamard matrix, in case
of required bit length of power of 2, or a Bernoulli
distribution $\Bern(0.5)$, for other bit lengths is used to generate the initial hash centers. Both approaches provide hash
centers reflecting distinct classes which are sufficiently far
apart with respect to the Hamming distance.
Each hash center $\vec{h}_c^{(0)} \text{, } c \in \{1, ..., C\}$ represents one
of the $C$ classes in the dataset. The index $0$ represents the initial state of the hash centers. These hash centers $\vec{h}_c^{(0)}$, are elements of $\mathbf{Y_h}$, which is one view of CCA loss $L_\mathrm{hash}$.
Opposed to \cite{yuan2020central} which keeps
hash centers constant during training, our approach
additionally performs an update after each training epoch with the goal
to better reflect semantic information in Hamming space.
Thus, the updated versions are able to better represent the class centers in Hamming
space more dynamically. Using this in the context of the DCCF loss results
in the target semantic hash centers being able to adapt to the semantic
information contained in the actual network outputs. Since a sigmoid activation is used in the hashing layer, the resulting hash
values $\vec{x_{h}}$ will be in the range between $[0, 1]$. These $\vec{x_{h}}$ are elements of $\mathbf{X_h}$ which is the second view of CCA loss $L_\mathrm{hash}$. The following proposed methods for
updating hash centers require them to be in the range of $[-1, 1]$. Therefore,
we apply a simple mapping, $f(\vec{x_{h}}) = 2  \vec{x_{h}} - 1 \text{ .}$ However, for valid further training the updated hash centers have to be
again binary values $\{0, 1\}$. Therefore, a final remapping to the required
binary space is performed.\par
\textbf{Single-labeled images:} After each training epoch, the hash values of each training image are
evaluated by a forward pass.  By using the
information of the class labels, the hashing outputs $\vec{x_{h}}$ can be grouped in sets
$\mathcal{G}_c^{(i)}$, each of which contains the hash values of images
associated with class $c$ at epoch $i$.  Now, for each group of hash codes, the
mean vector of all hashing values representing the same class is calculated as:
\begin{equation}
	\widetilde{\vec{h}}_c^{(i+1)} = \frac{1}{|\mathcal{G}_c^{(i)}|}
	\sum_{\vec{x_{h}}^{(i)} \in \mathcal{G}_c^{(i)}}  \vec{x_{h}}^{(i)} \text{ .}
	\label{eq:q}
\end{equation}
A subsequent thresholding of the obtained mean vector
$\widetilde{\vec{h}}_c^{(i+1)}$ results in the updated binary hash center
$\vec{h}_c^{(i+1)}$, with the thresholding function defined as 
\begin{equation}
	\tau(\vec{x}) = \begin{cases}
		1 \text{, if } x_k \geq 0 \\
		0 \text{, if } x_k <    0 \\
	\end{cases} \text{, for each element $x_k$ of $\vec{x}$.}
\end{equation}\par
\textbf{Multi-labeled images}: In this case, hash values can be associated with multiple
classes at the same time. Therefore, an extension of the hash center update given in eq.~\eqref{eq:q} is
introduced. The hash values are weighted with each update. Hashes are weighted lighter when containing more categories in their labels.
This is based on the assumption that with increasing category associations,
hash values are less able to represent a single category in Hamming space.
Therefore, a weighting factor $w_\mathrm{n} = \frac{1}{|l_\mathrm{n}|}$ for each of the $N$ hash
values is introduced based on the number of classes $|l_\mathrm{n}|$ in its corresponding
label $l_\mathrm{n}$. First, the hashes of the training images are evaluated by a
forward pass after weight update of the network in the current epoch. Then, the
obtained hash values are grouped in sets $\mathcal{G}_c^{(i)}$, each consisting
of those hashes which at least contain the respective class $c$ in their
corresponding label.  Note, that for multi-labeled data the resulting hash
value groups $\mathcal{G}_c^{(i)}$ are not distinct sets as in the previous
case of single-labeled data. This is because hash values labeled with multiple
categories $c$ at the same time are also contained in each corresponding group
$\mathcal{G}_c^{(i)}$. Now, the weights $w_\mathrm{n}$ are used to compute a weighted mean
of the hash values contained in each group $\mathcal{G}_c^{(i)}$ at epoch $i$:
\begin{equation}
	\widetilde{\vec{h}}_c^{(i+1)} = \frac{1}{|\mathcal{G}_c^{(i)}|}
	\sum_{\vec{x_{h}}^{(i)} \in \mathcal{G}_c^{(i)}} w_\mathrm{n} \, \vec{x_{h}}^{(i)} \text{ .}
\end{equation}
By applying a weighted mean, the contribution of each hash value is adapted
according to its reflecting semantic information about the class of a group.
The steps to perform a hash center update during DCSH training are summarized
in Algorithm \ref{al1}.% TODO: Add steps for binary space mappings ({0,1} -> {-1, 1} and reverse)
\begin{algorithm}[!htbp]
  \begin{algorithmic}[1]
    \REQUIRE Hash centers $\mathcal{H}^{(i)} = \{\vec{h}_c^{(i)}\}, c = 1, ...,
      C$ with $C$ classes, and output hashes
      $\{\vec{x_{h}}^{(i)}\} \in [0; 1]^n$ with $n = 1, ..., N$ from $N$
      training images at epoch $i$.
    \STATE For each class $c$, group all hash outputs associated with $c$ in their label $l_\mathrm{n}$: \\
      % \begin{equation*}
        $\mathcal{G}_c^{(i)} = \{\vec{x_{h}}^{(i)}: c \text{ in } l_\mathrm{n} \} \text{, for } c = 1, ..., C.$
      % \end{equation*}
    \STATE Calculate weights for each output hash based on number of classes in its label $|l_\mathrm{n}|$: \\
      % \begin{equation*}
        $w_\mathrm{n} = \frac{1}{|l_\mathrm{n}|}$.
      % \end{equation*}
    \STATE Calculate weighted mean of grouped hashing values: \\
      % \begin{equation*}
        $\widetilde{\vec{h}}_c^{(i+1)} = \frac{1}{|\mathcal{G}_c^{(i)}|}
          \sum\limits_{\vec{x_{h}}^{(i)} \in \mathcal{G}_c^{(i)}} w_\mathrm{n} \, \vec{x_{h}}^{(i)} \text{, for } c = 1, ..., C$.
      % \end{equation*}
    \STATE Create updated binary hash centers by thresholding: \\
      % \begin{equation*}
        $\vec{h}_c^{(i+1)} = \text{sign}(\widetilde{\vec{h}}_c^{(i+1)})$.
      % \end{equation*}
    \RETURN Updated hash centers $\mathcal{H}^{(i+1)}$ for epoch $i+1$.
  \end{algorithmic}
  \caption{DCSH hash center update}
  \label{al1}
\end{algorithm}
%\vspace{-1em}
\section{Experimental results}
\label{sec:expts}
In this paper, the single-labeled dataset CIFAR-10 \cite{krizhevsky2009learning}, and
multi-labeled datasets MS-COCO \cite{lin2014microsoft} and NUS-WIDE~\cite{chua2009nus} were used for evaluating the final retrieval performance. For
each dataset, the training and test curves and precision-recall (P-R) curves were plotted.
Furthermore, the retrieval performance was measured by calculating mean average precision (MAP) \cite{jiang2018deep}. The neural
network architecture used is ResNet-50 \cite{he2016deep}. We used a batchsize of $200$ in our experiments, with an initial learning rate and learning rate decay of $0.0003$ and $0.7$ every $10^{th}$ epoch for CIFAR-10 and $0.0008$ and $0.1$ for multi-labeled datasets which was determined by grid search. The optimizer used is Stochastic gradient descent. The regularizer $\alpha$ was chosen as shown in Eq. \eqref{eq:dcsh-lower_bound_combined}.\par
\begin{figure*}
	\centering
  %  	\begin{subfigure}{0.33\textwidth}
	%	\includegraphics[width=\textwidth]{figures/g-MNIST.pdf}%
	%	\caption{\small {Gallery MNIST}}
	%\end{subfigure}
	\begin{subfigure}{0.33\textwidth}
		\includegraphics[width=\textwidth]{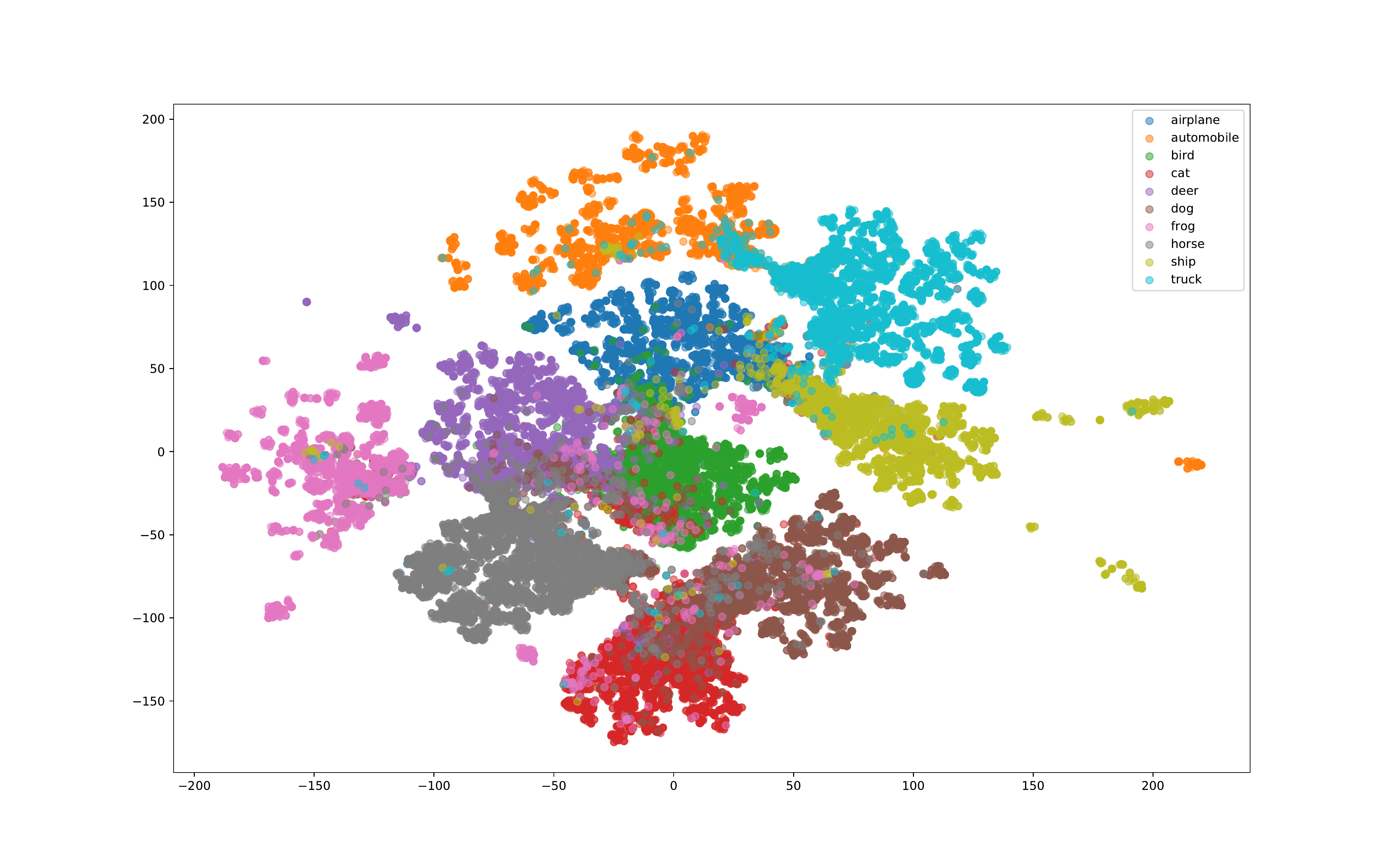}%
		\caption{\scriptsize{CIFAR-10 gallery features}}
	\end{subfigure}	
	\begin{subfigure}{0.33\textwidth}
        \includegraphics[width=\textwidth]{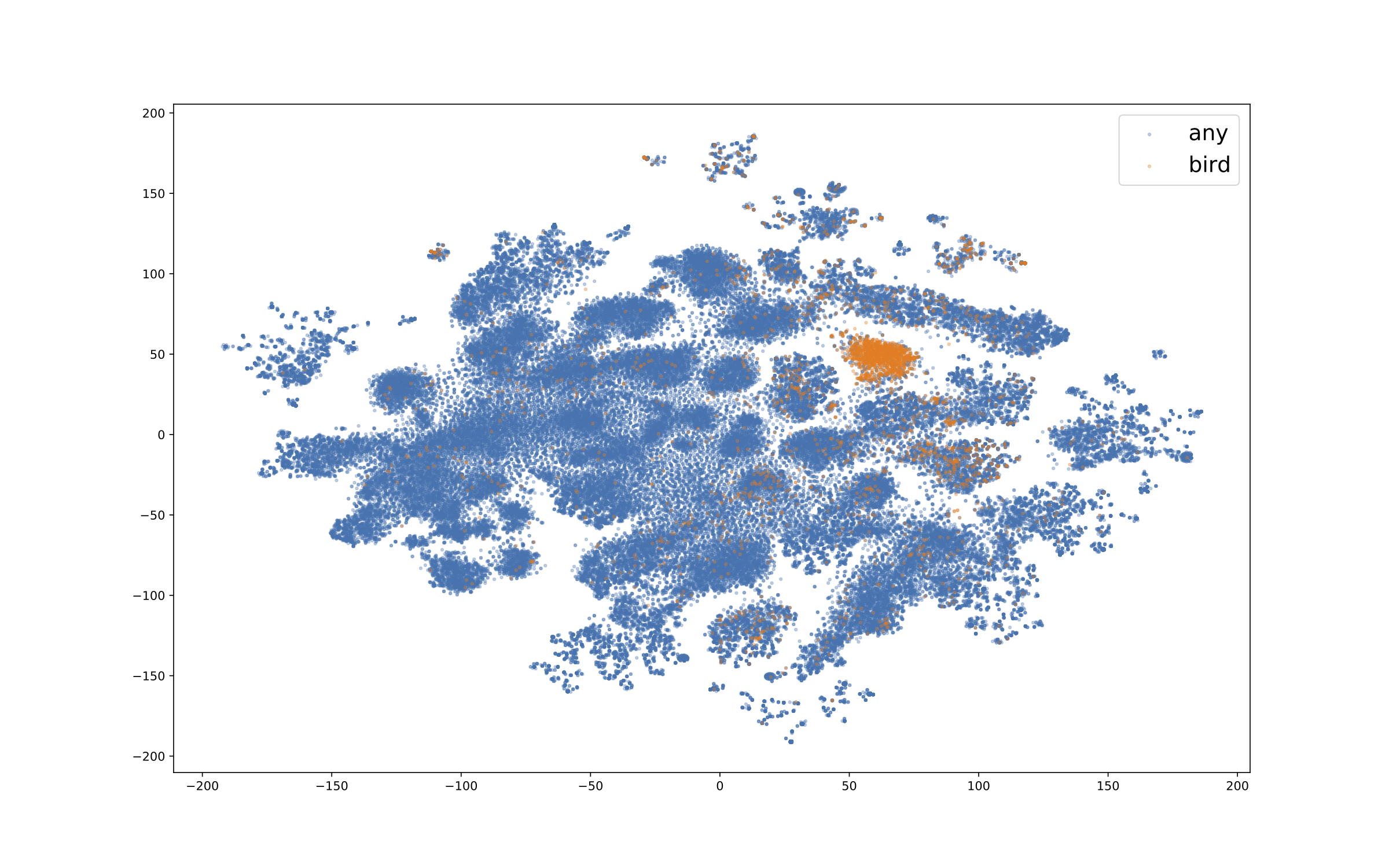}%
        \caption{ \scriptsize {MS-COCO bird category in orange}}
    \end{subfigure}
    \begin{subfigure}{0.33\textwidth}
		\includegraphics[width=\textwidth]{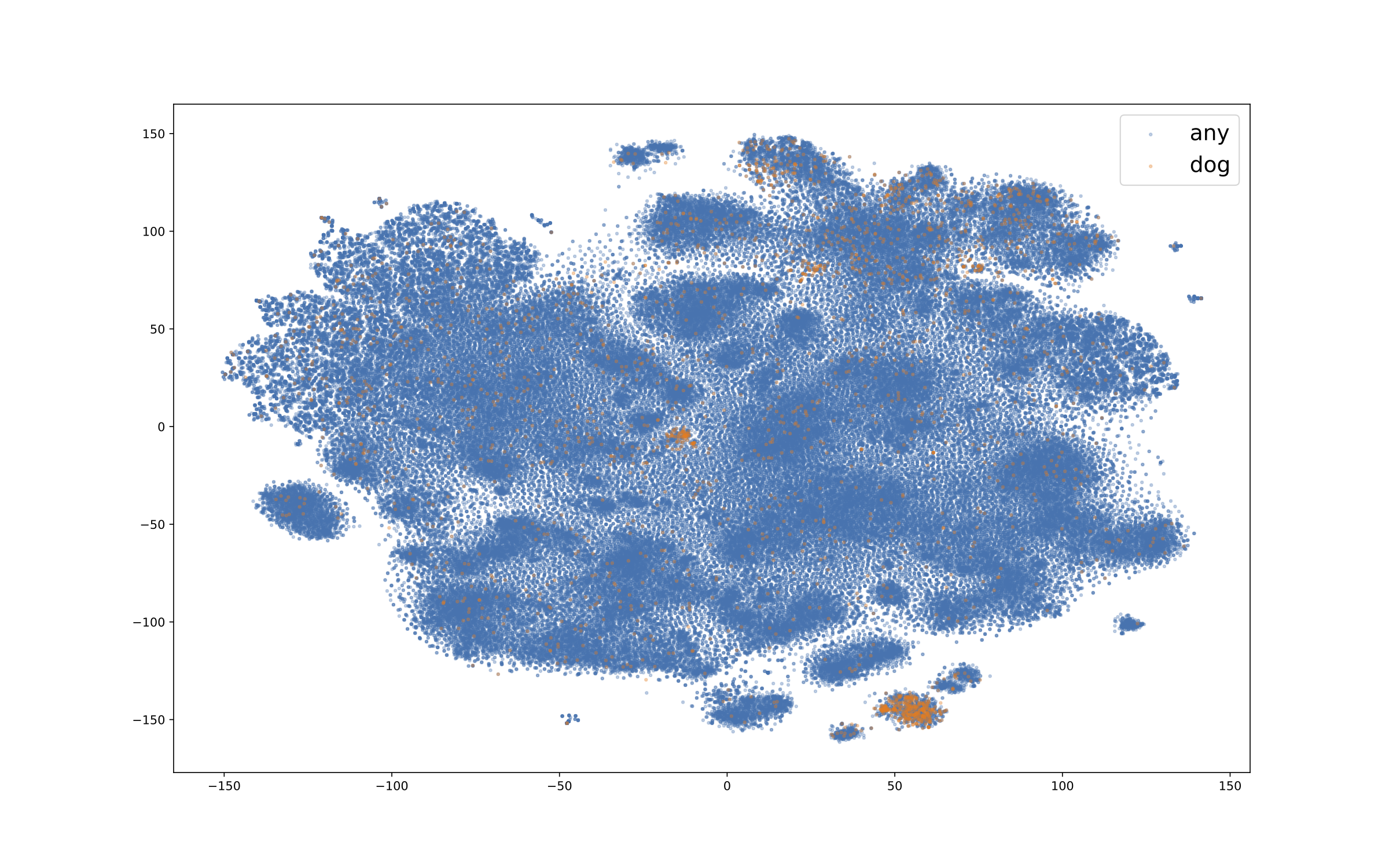}%
		\caption{\scriptsize {NUS-WIDE dog category in orange}}
		\end{subfigure}
	%\begin{subfigure}{0.49\textwidth}
	%	\includegraphics[width=\textwidth]{figures/NUSWIDE-flowers.png}%
	%	\caption{\small {NUS-WIDE Flowers Category}}
	%		\end{subfigure}
	%	\begin{subfigure}{0.33\textwidth}
	  %  \includegraphics[width=\textwidth]{figures/NUSWIDE-dog.png}%
	 %   \caption{\small {NUS-WIDE Dog Category}}
	 %   \end{subfigure}
	\label{fig:results-tsne_of_bin_hashes_mnist_gallery}
	\caption{{T-SNE of $64$ dimensional features for a) gallery images of CIFAR-10, b) MS-COCO highlighting bird category, and c) NUS-WIDE highlighting dog category. } }
	\label{fig:tsne}
\end{figure*}
%\begin{table}[htbp!]
%	\begin{center}
%			\caption{Mean average precision for MNIST}
%		\begin{tabular}{ l | c | c | c | c }
%			\toprule
%			\multirow{2}{*}{Method}           & \multicolumn{4}{c}{MNIST (mAP@5000)} \\
%			\cline{2-5}
%			% & 12 bits & 16 bits & 32 bits & 64 bits \\
%			& 16 bits         & 32 bits         & 48 bits         & 64 bits         \\
%			\midrule
%			% & 12 bits & 16 bits & 32 bits & 64 bits \\
%			\textbf{DCSH (Ours)}              & \textbf{0.9909} & \textbf{0.9883} & \textbf{0.9905} & \textbf{0.9918} \\
%			\midrule
%			DRSCH \cite{zhang2015bit}         & 0.969           & 0.979           & 0.979           & 0.981           \\
%			DSCH \cite{}                      & 0.965           & 0.972           & 0.974           & 0.978           \\
%			DSRH \cite{zhao2015deep}          & 0.965           & 0.972           & 0.975           & 0.978           \\
%			KSH-CNN \cite{liu2012supervised}  & 0.839           & 0.885           & 0.894           & 0.897           \\
%			MLH-CNN \cite{norouzi2011minimal} & 0.710           & 0.781           & 0.807           & 0.809           \\
%			\bottomrule
%		\end{tabular}
%		\label{tab:results-mnist_map}
%	\end{center}
%\end{table}
\textbf{CIFAR-10} \cite{krizhevsky2009learning} is a labeled
subset of tiny images \cite{torralba200880} with $10$ classes. Each class is equally represented by $6000$ images making a total of $60,000$ images. In order to better reflect the requirements of an image
retrieval application, the following setup is used in accordance to the
experimental setup in DDSH \cite{jiang2018deep}. From the total
$60,000$ images, $1000$ images are used as query images by sampling $100$ random images from each of the ten classes of the dataset. The remaining $59,000$ images
constitute the gallery set against which retrievals are performed.
Furthermore, $5000$ training images are randomly sampled from the gallery set such
that each class is again equally represented with $500$ images.
Thus, the training set is a labeled subset of the gallery set. The experiments were run for
$25$ epochs. The training and test loss during optimization for binary codes of $32$ bits is
depicted in Fig. \ref{fig:loss_pr} (a). (The loss curves and P-R curves for other bits are given in the supplementary material.) The theoretical lower bound of $-40$ as mentioned in Eq. \eqref{eq:dcsh-lower_bound_combined} is reached during training.
The distribution of the resulting binary hashes in Hamming space is visualized
in Fig. \ref{fig:tsne} (a) by using t-SNE for the gallery images. Binary features associated with the same class are located close to each other
while having good separability to binary features of other categories.
Furthermore, semantically related categories like 'automobile', 'airplane'
or 'truck' are located next to each other in Hamming space, while categories
like 'dog' or 'cat' are located on the other side.
The MAP@5000 was measured and the
results are outlined in Table~\ref{tab:results-cifar10_map}. It can be seen
that the proposed DCSH method clearly outperforms existing state-of-the-art approaches. Furthermore, P-R curves for $32$ bits was plotted in Fig. \ref{fig:loss_pr} (d). P-R curves were generated by retrieving all relevant images in gallery whereas MAP values in Table \ref{tab:results-cifar10_map} denotes MAP@5000 in which $5000$ returned images was considered. Here, the
precision is high for a large range of recall values. The precision
only starts to decrease for high values of recall. Therefore, the majority
of relevant images with respect to a given query image are likely to be
returned within the top retrievals.\par
\begin{figure*}
	\centering
	%\begin{subfigure}{0.245\textwidth}
	%	\includegraphics[width=\textwidth]{figures/MNIST_32.pdf}%
	%	\caption{\small {MNIST loss curve }}
	%\end{subfigure}

	\begin{subfigure}{0.24\textwidth}
	\includegraphics[width=0.7\textwidth,  height=0.47\textwidth]{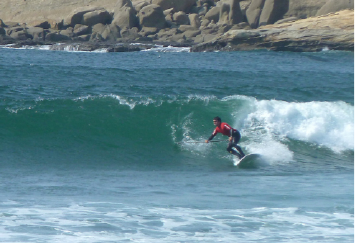}%
	\caption{\scriptsize {Query image 1}}
\end{subfigure}	
\begin{subfigure}{0.24\textwidth}
	\includegraphics[width=0.7\textwidth,  height=0.47\textwidth]{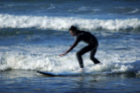}%
	\caption{\scriptsize {Retrieval result 1}}
\end{subfigure}
\begin{subfigure}{0.24\textwidth}
	\includegraphics[width=0.7\textwidth,  height=0.47\textwidth]{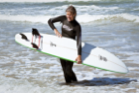}%
	\caption{ \scriptsize{Retrieval result 2}}
\end{subfigure}
\begin{subfigure}{0.24\textwidth}
	\includegraphics[width=0.7\textwidth,  height=0.47\textwidth]{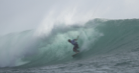}%
	\caption{\scriptsize {Retrieval result 3}}
\end{subfigure}
\caption{{Query image and top $3$ retrievals on MS-COCO, all contain category labels ’surfboard’ and ’person’.
}}
\label{ret:surf}

\end{figure*}

\begin{table}[t]
	\begin{center}
		\caption{{MAP@5000 for CIFAR-10.}}
		\begin{tabular}{ l  c  c  c  c }
			\toprule
		Method	& 12 bits        & 24 bits        & 32 bits        & 48 bits        \\
			\midrule
			\textbf{DCSH (Ours)}           & \textbf{0.863} & \textbf{0.898} & \textbf{0.902} & \textbf{0.911} \\ % Results with ResNet50
			% DCCH (Ours)                   & 0.863   & 0.890   & 0.887   & 0.901   \\ % Old results with ResNet34
		%	\midrule
		
			% Results from DCCH paper
			DCCH \cite{jose2020optimized}  & 0.794          & 0.830          & 0.841          & 0.851          \\
			DDSH \cite{jiang2018deep}      & 0.769          & 0.829          & 0.815          & 0.819          \\
			DSDH \cite{li2017deep}         & 0.740          & 0.786          & 0.801          & 0.820          \\
						LDH \cite{hong2020image}       & -              & -              & 0.784          & 0.792          \\
			DTSH \cite{wang2016deep}       & 0.710          & 0.750          & 0.765          & 0.774          \\

			%	\midrule
			DPSH \cite{li2015feature}      & 0.713          & 0.727          & 0.744          & 0.757                  \\

		%	\midrule
		%	NDH \cite{ma2017nonlinear}     & 0.562          & 0.640          & 0.652          & 0.677          \\
		%	COSDISH \cite{kang2016column}  & 0.609          & 0.683          & 0.696          & 0.716          \\
		%	SDH \cite{shen2015supervised}  & 0.520          & 0.646          & 0.658          & 0.669          \\
		%	FastH \cite{lin2014fast}       & 0.629          & 0.673          & 0.687          & 0.713          \\
		%	LFH \cite{zhang2014supervised} & 0.401          & 0.605          & 0.657          & 0.700          \\
			\bottomrule
		\end{tabular}
		\label{tab:results-cifar10_map}
	\end{center}
\end{table}
\begin{figure*}
	\centering
	%\begin{subfigure}{0.245\textwidth}
	%	\includegraphics[width=\textwidth]{figures/MNIST_32.pdf}%
	%	\caption{\small {MNIST loss curve }}
	%\end{subfigure}
	\begin{subfigure}{0.33\textwidth}
		\includegraphics[width=0.9\textwidth,  height=0.56\textwidth]{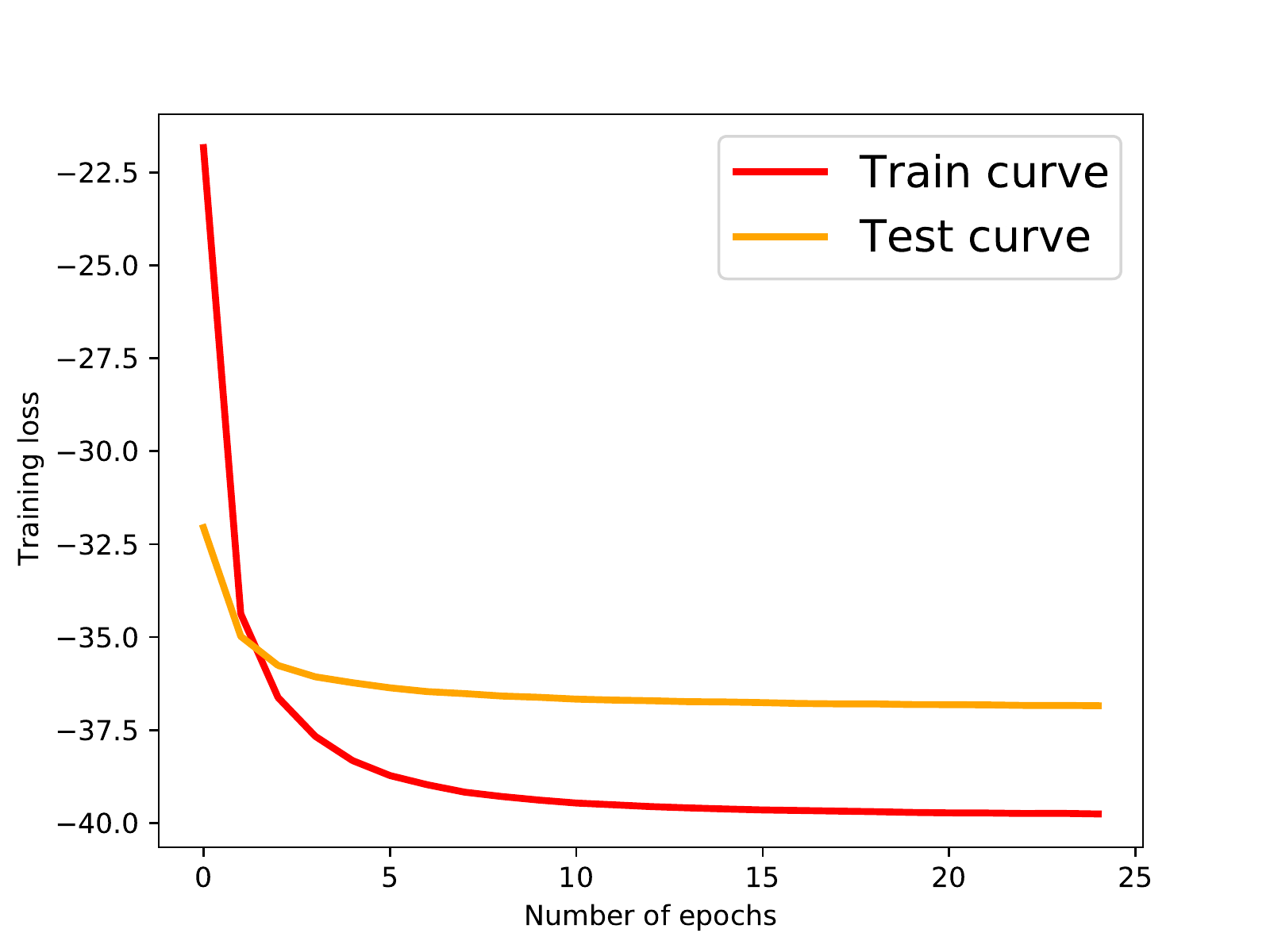}%
		\caption{\scriptsize {CIFAR-10 loss curve}}
	\end{subfigure}	
	\begin{subfigure}{0.33\textwidth}
		\includegraphics[width=0.9\textwidth,  height=0.56\textwidth]{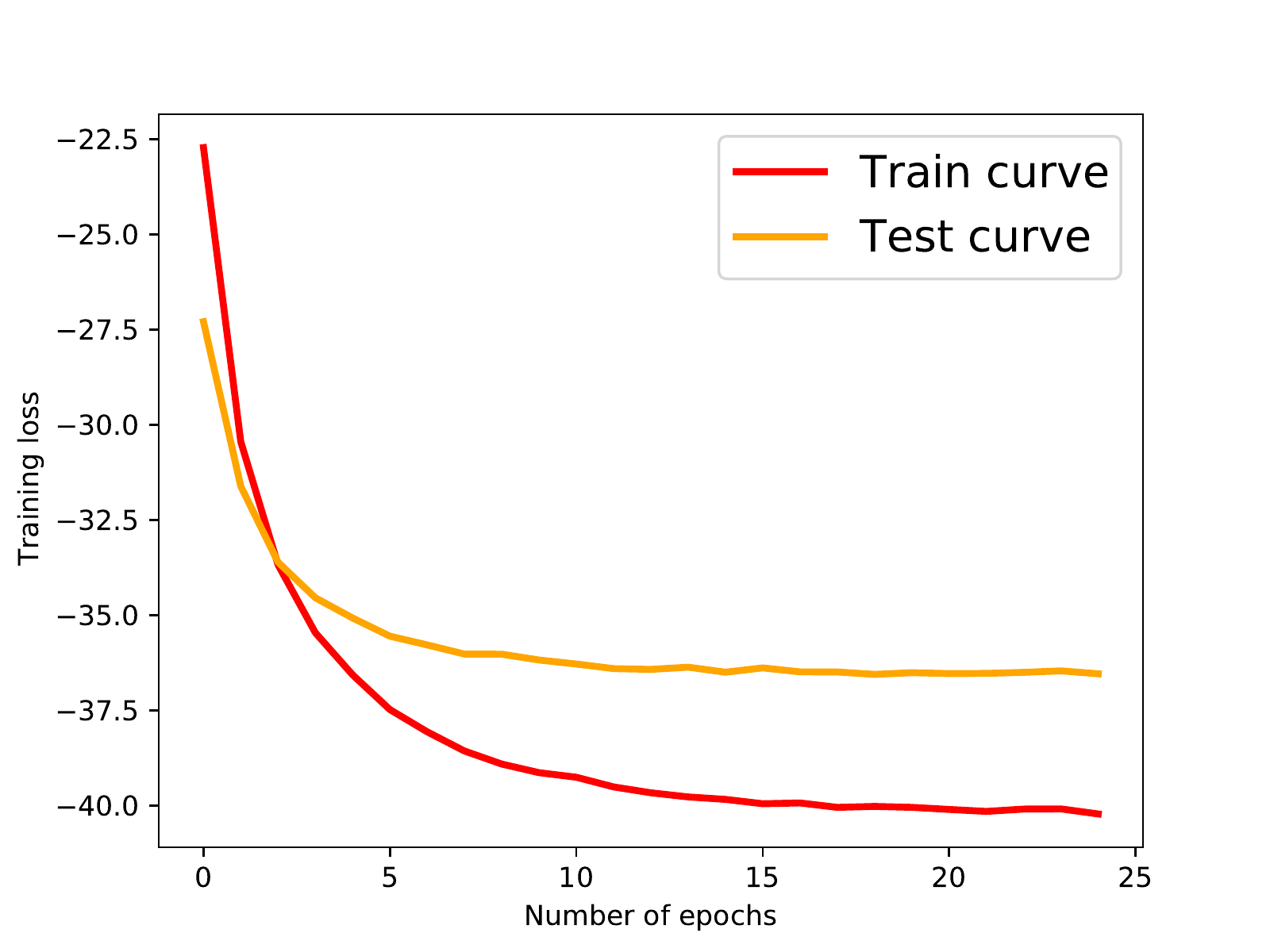}%
		\caption{ \scriptsize {MS-COCO loss curve}}
	\end{subfigure}
	\begin{subfigure}{0.33\textwidth}
		\includegraphics[width=0.9\textwidth,  height=0.56\textwidth]{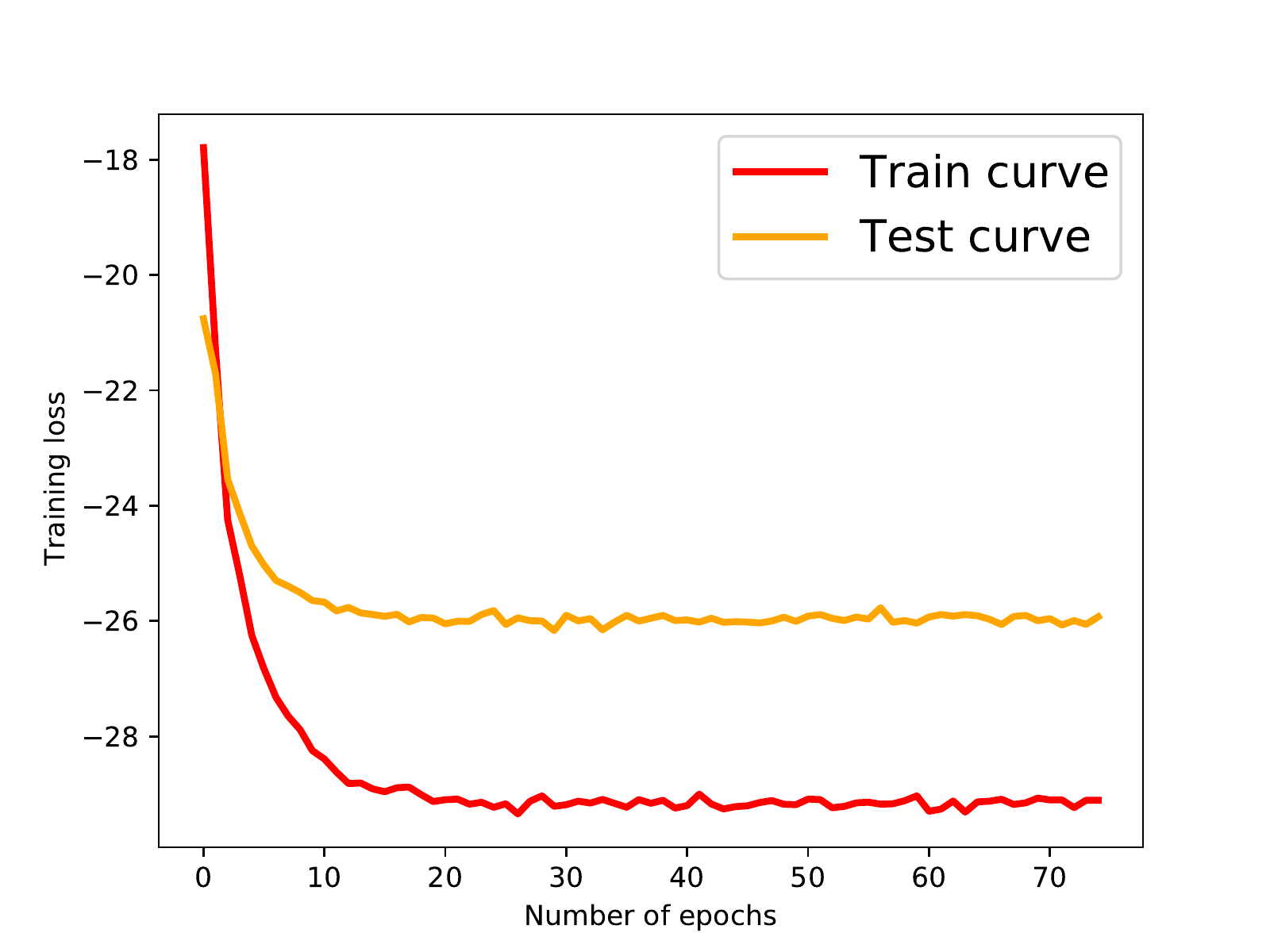}%
		\caption{\scriptsize {NUS-WIDE loss curve}}
	\end{subfigure}
	%\begin{subfigure}{0.245\textwidth}
	%	\includegraphics[width=\textwidth]{figures/precision_recall_curve_MNIST_bits_32.png}%
	%	\caption{\small {MNIST p-r curve}}
	%\end{subfigure}
	\begin{subfigure}{0.33\textwidth}
		\includegraphics[width=0.9\textwidth,  height=0.56\textwidth]{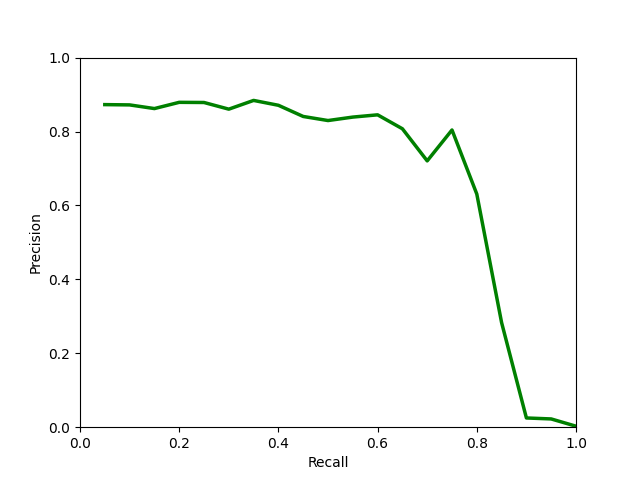}%
		\caption{\scriptsize {CIFAR-10 P-R curve }}
	\end{subfigure}
	\begin{subfigure}{0.33\textwidth}
	\includegraphics[width=0.9\textwidth , height=0.56\textwidth]{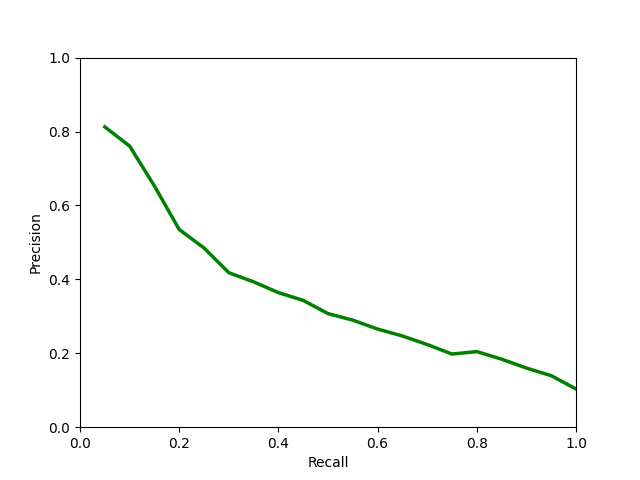}%
	\caption{\scriptsize {MS-COCO P-R curve }}
    \end{subfigure}
	\begin{subfigure}{0.33\textwidth}
	\includegraphics[width=0.9\textwidth, height=0.56\textwidth]{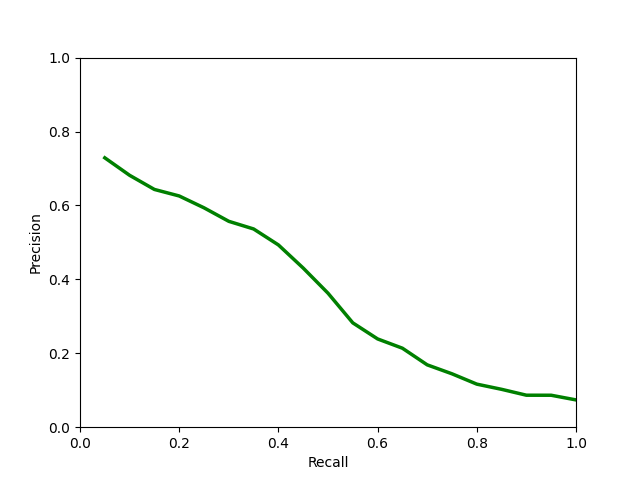}%
	\caption{\scriptsize {NUS-WIDE P-R curve }}
    \end{subfigure}	
	\caption{First row indicates the training and test loss for all the three datasets for $32$ bits. The training loss reaches the lower bound of $-40$ for CIFAR-10. For multi-labeled, the lower bound of $-62$ could not be reached as the correlation coefficients will never reach the maximum value. Second row indicates the P-R curves for all the three datasets for $32$ bits.}
	\label{fig:loss_pr}
\end{figure*}
\textbf{MS-COCO} \cite{lin2014microsoft} is a
multi-labeled dataset where each image is associated with several of $80$
distinct object categories. The used data split for image retrieval evaluation
is based on the setup described in HashNet \cite{cao2017hashnet}. Combining training and validation images and subsequent discard of all images
without category information results in a total number of $122,218$ images. $5,000$ query images are randomly sampled from these obtained images, leaving the
remaining $117,218$ images as the gallery set. In addition, $10,000$ images are
randomly sampled from the gallery set for training. The model was trained for $25$ epochs.
The training and test loss, values during training for $32$ bits is given in
Fig. \ref{fig:loss_pr} (b). Theoretically, the lower bound of the loss function
is $- 2(B - 1)$ with $B$ number of bits, as discussed in Eq. \eqref{eq:dcsh-lower_bound_combined}. However, it can be seen that this lower bound is not reached. This is because
a multi-labeled dataset cannot be fully class-wise clustered as each image may belong to several categories. Therefore,
the correlation coefficients of the underlying loss are not able to reach
the maximum. The training reaches a lower bound indicating formation of an optimum feature space. Since MS-COCO is a multi-labeled dataset, it is not
feasible to depict all class associations for the data points at once by using
different colors for each categories. Instead, single categories have
been selected to be highlighted against the remaining categories of the
dataset. Clusters of binary hashes are formed in the
feature space. There are cases where these clusters represent
a unique category, as clearly shown in the example of 'bird' depicted
in Fig. \ref{fig:tsne} (b).
The retrieval performance of DCSH for MS-COCO in terms
of MAP is given in Table~\ref{tab:results-mscoco_map}. Results are compared to state-of-the-art approaches. It can be seen that CSQ is having same MAP for $64$
bits and in all other cases, the proposed method outperforms other retrieval frameworks. CSQ did not provide the MAP for
the case of $48$ bits in \cite{yuan2020central}. The
P-R curves for $32$ bits are shown in Fig.
\ref{fig:loss_pr} (e). The top $3$ retrieval results for a query image with labels 'surfboard' and 'person' are shown in Fig. \ref{ret:surf}.\par
%\vspace{-em}
\begin{table}[t]
	\begin{center}
		\caption{{MAP@5000 for MS-COCO.}}
		\begin{tabular}{ l  c  c  c  c }
			\toprule
		Method	& 16 bits        & 32 bits        & 48 bits        & 64 bits        \\
			\midrule
			\textbf{DCSH (Ours)}             & \textbf{0.805} & \textbf{0.847} & \textbf{0.859} & \textbf{0.861}          \\
		%	\midrule
			CSQ \cite{yuan2020central}       & 0.796          & 0.838          & -              & \textbf{0.861} \\
		%	\midrule
			% Results from DCCH paper (Jose)
			DCCH \cite{jose2020optimized}    & 0.659          & 0.729          & 0.731          & 0.739          \\
			HashNet \cite{cao2017hashnet}    & 0.687          & 0.718          & 0.730          & 0.736          \\
			DHN \cite{zhu2016deep}           & 0.677          & 0.701          & 0.695          & 0.694          \\
			DNNH \cite{lai2015simultaneous}  & 0.593          & 0.603          & 0.604          & 0.610          \\
			CNNH \cite{xia2014supervised}    & 0.564          & 0.574          & 0.571          & 0.567          \\
		%	\midrule
		%	SDH \cite{shen2015supervised}    & 0.555          & 0.564          & 0.572          & 0.580          \\
		%	KSH \cite{liu2012supervised}     & 0.521          & 0.534          & 0.534          & 0.536          \\
		%	ITQ-CCA \cite{gong2012iterative} & 0.566          & 0.562          & 0.530          & 0.502          \\
		%	\midrule
		%	ITQ \cite{gong2012iterative}     & 0.582          & 0.624          & 0.646          & 0.657          \\
		%	BRE \cite{kulis2009learning}     & 0.592          & 0.622          & 0.630          & 0.634          \\
		%	SH \cite{weiss2008spectral}      & 0.495          & 0.507          & 0.510          & 0.510          \\
		%	\midrule
		%	LSH \cite{gionis1999similarity}  & 0.459          & 0.486          & 0.544          & 0.585          \\
			%%%
			\bottomrule
		\end{tabular}
		\label{tab:results-mscoco_map}
	\end{center}
\end{table}
\begin{figure*}[t]
	\centering
	%\begin{subfigure}{0.245\textwidth}
	%	\includegraphics[width=\textwidth]{figures/MNIST_32.pdf}%
	%	\caption{\small {MNIST loss curve }}
	%\end{subfigure}
	\begin{subfigure}{0.24\textwidth}
		\includegraphics[width=0.75\textwidth,  height=0.67\textwidth]{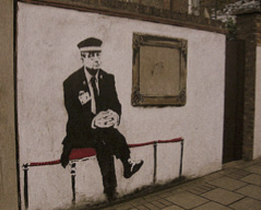}%
		\caption{ \scriptsize{Query image 1}}
	\end{subfigure}	
	\begin{subfigure}{0.24\textwidth}
		\includegraphics[width=0.75\textwidth,  height=0.67\textwidth]{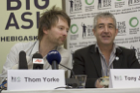}%
	\caption{\scriptsize {Retrieval result 1}}
	\end{subfigure}
	\begin{subfigure}{0.24\textwidth}
		\includegraphics[width=0.6\textwidth,  height=0.67\textwidth]{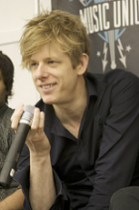}%
		\caption{ \scriptsize{Retrieval result 2}}
	\end{subfigure}
	\begin{subfigure}{0.24\textwidth}
		\includegraphics[width=0.6\textwidth,  height=0.67\textwidth]{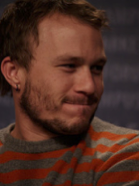}%
	\caption{ \scriptsize{Retrieval result 3}}
	\end{subfigure}
	\caption{{Query image and top $3$ retrievals on NUS-WIDE, all contain category label 'person'.}}
	\label{ret:per}
\end{figure*}
\textbf{NUS-WIDE} was introduced by Chua et al. in \cite{chua2009nus}
containing $269,648$ web images in total which are associated with $5,018$ unique
tags from Flickr. It is a multi-labeled dataset in which each image is associated with one or more of $81$ concepts. The experimental setup for NUS-WIDE in DDSH \cite{jiang2018deep} is used here. Only the images belonging to the $10$ most frequent
concepts are used. This results in a total number of $186,577$ images, from which
$1,867$ query images are randomly sampled \cite{jiang2018deep}. The remaining
$184,710$ images constitute the gallery set, and $5,000$ images are randomly
sampled from it for training \cite{jiang2018deep}. The network was trained for 75 epochs. Fig.
\ref{fig:loss_pr} (c) shows the training and test loss
during training for $32$ bits. The theoretical lower-bound in this case is $-2(B - 1)$ as well, and cannot be reached as explained in the case of MS-COCO. In Fig.
\ref{fig:tsne} (c),
orange color indicates the 'dog' category. It can be seen that there is
clearly a dominant area in the feature space. Therefore, a nearest neighbor search would result in
retrieved images very likely being associated with the similar semantics.
MAP has been measured for the proposed
approach and the numbers were compared to
state-of-the-art approaches. The results are outlined in Table ~\ref{tab:results-nuswide_map}. DCSH is
mainly compared to CSQ \cite{yuan2020central} and DCCH
\cite{jose2020optimized}. The proposed DCSH method outperforms other approaches. Note that CSQ did not list the MAP for $12$ and $24$ bits in
\cite{yuan2020central}. The MAP for $16$ bits is given in
\cite{yuan2020central} and is $0.810$, for which DCSH gave a
value of $0.828.$ For further evaluation, the precision-recall curves were plotted.
For DCSH for $32$ bits P-R curve is given in Fig.~\ref{fig:loss_pr} (f). The precision drops as recall increases as many wrong retrievals occur till we retrieve all relevant images. An example query image and the top $3$ retrieval results for category 'person' are shown in Fig.~\ref{ret:per}.
%\vspace{-2em}
\begin{table}[t]
	\caption{{MAP@5000 for NUS-WIDE.}}
	\begin{center}
		\begin{tabular}{ l  c  c  c  c }
\midrule
		Method	& 12 bits        & 24 bits        & 32 bits        & 48 bits        \\
			\midrule
			\textbf{DCSH (Ours)}           & \textbf{0.823} & \textbf{0.833} & \textbf{0.841} & \textbf{0.857} \\
		%	\midrule
		DPSH \cite{li2015feature}      & 0.794          & 0.822          & 0.838          & 0.851          \\
					DCCH \cite{jose2020optimized}  & 0.782          & 0.814          & 0.825          & 0.834          \\
			CSQ \cite{yuan2020central}     & -              & -              & 0.825          & 0.832          \\

		%	\midrule
					
					DSDH \cite{li2017deep}         & 0.776          & 0.808          & 0.820          & 0.829
\\			DDSH \cite{jiang2018deep}      & 0.791          & 0.815          & 0.821          & 0.827          \\
          
			DTSH \cite{wang2016deep}       & 0.773          & 0.808          & 0.812          & 0.814          \\
			LDH \cite{hong2020image}       & 0.769              & 0.789             & 0.787          & 0.803          \\

		%	\midrule
		%	NDH \cite{ma2017nonlinear}     & 0.702          & 0.735          & 0.745          & 0.745          \\
		%	COSDISH \cite{kang2016column}  & 0.730          & 0.764          & 0.787          & 0.799          \\
		%	SDH \cite{shen2015supervised}  & 0.739          & 0.762          & 0.770          & 0.772          \\
		%	FastH \cite{lin2014fast}       & 0.741          & 0.783          & 0.795          & 0.809          \\
		%	LFH \cite{zhang2014supervised} & 0.705          & 0.759          & 0.778          & 0.794          \\
			\bottomrule
		\end{tabular}
		\label{tab:results-nuswide_map}
	\end{center}
\end{table}
\section{Conclusions}
\label{sec:conc}
In this paper, we have proposed an approach for learning efficient hash codes for image retrieval. The neural network is trained using a loss function in such a way that the correlation between the hash codes and hash centers is maximized. Canonical Correlation Analysis (CCA) is utilized in the loss formulation, and hash codes and hash centers are chosen as the two views of CCA. The network is also trained using the classification loss as well, which maximizes the correlation between category labels and classification scores. The hash centers are then dynamically updated so that the hash centers adapt to the changes in feature space. The experimental results on both single-labeled and multi-labeled datasets substantiates the generation of an optimized feature space with minimum intra-class scatter and maximum inter-class scatter. This is in fact possible due to the inherent equivalence between Linear Discriminant Analysis~\cite{lda} and CCA as proven in~\cite{bartlett1938further}. As a future research direction, more effective representations of individual categories could be explored for multi-labeled datasets, which in turn could lead to better retrieval results.
%A future research direction would be the extension of experiments to video retrieval. 
% % TODO: Change that part. I don't like it.
% After applying Deep Central Similarity Hashing (DCSH) to a dataset, binary
% codes of the retrieval database and query images can be obtained for efficient
% image retrieval. Given an example query image, the generated binary hash value
% representing it can be used to perform a nearest neighbor search to the binary
% codes of the retrieval database. The images of the gallery set corresponding to
% the closest hash values to the query with respect to the Hamming distance are
% then the resulting retrievals.

{\small
\bibliographystyle{ieee_fullname}
\bibliography{egbib}
}

\end{document}

	%%%%%%%%% TITLE
	%\title{Realistic Video Face Retargeting Using Conditional Generative Adversarial Networks}
	% \title{Animating Realistic Dynamic Facial Textures from a Single Image using GANs}
	\title{Learning Classical Linear Discriminant Analysis Metric Using a Siamese Neural Network- Supplementary Material}
	
%	\author[1,3,4]{Kyle Olszewski\thanks{olszewski.kyle@gmail.com (equal contribution)}}
%	\author[1]{Zimo Li\thanks{zimoli@usc.edu (equal contribution)}}
%	\author[1]{Chao Yang \thanks{harryyang.hk@gmail.com (equal contribution)}}
%	\author[1]{Yi Zhou\thanks{zhou859@usc.edu}}
%	\author[1,3]{Ronald Yu\thanks{ronaldyu@usc.edu}}
%	\author[1]{Zeng Huang\thanks{zenghuan@usc.edu}}
%	\author[1]{Sitao Xiang\thanks{sitaoxia@usc.edu}}
%	\author[1,3]{Shunsuke Saito\thanks{shunsuke.saito16@gmail.com}}
%	\author[2]{Pushmeet Kohli\thanks{pushmeet@google.com, project conducted while at MSR}}
%	\author[1,3,4]{Hao Li\thanks{hao@hao-li.com}}
%	\affil[1]{University of Southern California}
%	\affil[2]{DeepMind}
%	\affil[3]{Pinscreen}
%	\affil[4]{USC Institute for Creative Technologies}
	%\title{Animating Realistic Dynamic Facial Textures using Generative Adversarial Networks}
	
	%% \author{First Author\\
	%% Institution1\\
	%% Institution1 address\\
	%% {\tt\small firstauthor@i1.org}
	%% % For a paper whose authors are all at the same institution,
	%% % omit the following lines up until the closing ``}''.
	%% % Additional authors and addresses can be added with ``\and'',
	%% % just like the second author.
	%% % To save space, use either the email address or home page, not both
	%% \and
	%% Second Author\\
	%% Institution2\\
	%% First line of institution2 address\\
	%% {\tt\small secondauthor@i2.org}
	%% }
	
	\maketitle
	\thispagestyle{empty}

\section{Network architecture}
The network architecture used for training the Siamese Neural Network (SNN) is described in Table~\ref{network}. The Convolutional Neural Network (CNN) architecture is shown below since the SNN consists of two parallel CNNs with shared weights.
\begin{table*}[ht]
	\centering
	\begin{tabular}{|l| l| l|}
		\hline
		MNIST                                   & CIFAR-10                                & STL-10                                  \\ \hline\hline
		Input $1\times28\times28$               & Input $3\times32\times32$               & Input $3\times96\times96$               \\
		$3\times3$ conv, $32,$ BN-ReLU, (0.3)   & $3\times3$ conv, $64,$ BN-ReLU, (0.3)   & $3\times3$ conv, $64,$ BN-ReLU,  (0.5)  \\
		$3\times3$ conv, $32,$ BN-ReLU          & $3\times3$ conv, $64,$ BN-ReLU          & $3\times3$ conv, $64,$ BN-ReLU          \\
		$2\times2$ max-pool                     & $2\times2$ max-pool                     & $2\times2$ max-pool                     \\
		$3\times3$ conv, $64,$ BN-ReLU, (0.4)   & $3\times3$ conv, $128,$ BN-ReLU,  (0.4) & $3\times3$ conv, $64,$ BN-ReLU, (0.5)   \\
		$3\times3$ conv, $64,$ BN-ReLU          & $3\times3$ conv, $128,$ BN-ReLU         & $3\times3$ conv, $128,$ BN-ReLU         \\
		$2\times2$ max-pool                     & $2\times2$ max-pool                     & $2\times2$ max-pool                     \\
		$3\times3$ conv, $128,$ BN-ReLU,  (0.4) & $3\times3$ conv, $256,$ BN-ReLU,  (0.4) & $3\times3$ conv, $128,$ BN-ReLU, (0.5)  \\
		$3\times3$ conv, $128,$ BN-ReLU,  (0.4) & $3\times3$ conv, $256,$ BN-ReLU,  (0.4) & $3\times3$ conv, $256,$ BN-ReLU, (0.5)  \\
		$2\times2$ max-pool                     & $3\times3$ conv, $256,$ BN-ReLU         & $3\times3$ conv, $256,$ BN-ReLU         \\
		Linear $(128-128)$                      & $2\times2$ max-pool                     & $2\times2$ max-pool                     \\
		Linear $(128-9)$                        & $3\times3$ conv, $512,$ BN-ReLU,  (0.4) & $3\times3$ conv, $256,$ BN-ReLU,  (0.5) \\
		KL-divergence                           & $3\times3$ conv, $512,$ BN-ReLU,  (0.4) & $3\times3$ conv, $512,$ BN-ReLU,  (0.5) \\
		LDA-Loss                                & $3\times3$ conv, $512,$ BN-ReLU         & $3\times3$ kernel, $512,$ BN-ReLU       \\
		& $2\times2$ max-pool                     & $2\times2$ max-pool                     \\
		& $3\times3$ conv, $512,$ BN-ReLU, (0.4)  & $3\times3$ conv, $512,$ BN-ReLU,  (0.5) \\
		& $3\times3$ conv, $512,$ BN-ReLU,  (0.5) & $3\times3$ conv, $512,$ BN-ReLU,  (0.5) \\
		& $3\times3$ conv, $512,$ BN-ReLU         & $3\times3$ conv, $512,$ BN-ReLU         \\
		& $2\times2$ max-pool                     & $2\times2$ max-pool                     \\
		& Linear $(512-512)$                      & Linear $(512*3*3-512)$                  \\
		& Linear $(512-9)$                        & Linear $(512-9)$                        \\
		& KL-divergence                           & KL-divergence                           \\
		& LDA-Loss                                & LDA-Loss                                \\ 
		\hline
		
	\end{tabular}
	
	\caption{\footnotesize Network architecture used in our experiments for the three datasets. Conv - convolutional kernel, BN - Batch Normalization, ReLU - Rectified Linear Unit, \% of introduced dropout is shown in brackets.}
	\label{network}
\end{table*}

\section{Projection of feature vectors along the eigenvector directions at the end of training}
The projection of feature vectors at the end of training is shown in Figure \ref{feature_proj}. One can observe that the within-class scatter has reduced and between-class scatter has increased resulting in a class projection such that the classes are linearly separable along the projected discriminant directions.

\begin{figure*}
	\centering

	\begin{subfigure}{0.33\textwidth}
		\includegraphics[width=0.9\textwidth,  height=0.56\textwidth]{Imgs-2/1.png}%
		\caption{\scriptsize {Eigenvector 1}}
	\end{subfigure}	
	\begin{subfigure}{0.33\textwidth}
		\includegraphics[width=0.9\textwidth,  height=0.56\textwidth]{Imgs-2/2.png}%		\caption{ \scriptsize {Eigenvector 2}}
	\end{subfigure}
	\begin{subfigure}{0.33\textwidth}
		\includegraphics[width=0.9\textwidth,  height=0.56\textwidth]{Imgs-2/3.png}%
		\caption{\scriptsize {Eigenvector 3}}
	\end{subfigure}
	%\begin{subfigure}{0.245\textwidth}
	%	\includegraphics[width=\textwidth]{figures/precision_recall_curve_MNIST_bits_32.png}%
	%	\caption{\small {MNIST p-r curve}}
	%\end{subfigure}
	\begin{subfigure}{0.33\textwidth}
		\includegraphics[width=0.9\textwidth,  height=0.56\textwidth]{Imgs-2/4.png}%
		\caption{\scriptsize {Eigenvector 4 }}
	\end{subfigure}
	\begin{subfigure}{0.33\textwidth}
		\includegraphics[width=0.9\textwidth , height=0.56\textwidth]{Imgs-2/5.png}%
		\caption{\scriptsize {Eigenvector 5 }}
	\end{subfigure}
	\begin{subfigure}{0.33\textwidth}
		\includegraphics[width=0.9\textwidth, height=0.56\textwidth]{Imgs-2/6.png}%
		\caption{\scriptsize {Eigenvector 6 }}
	\end{subfigure}

	\begin{subfigure}{0.33\textwidth}
	\includegraphics[width=0.9\textwidth,  height=0.56\textwidth]{Imgs-2/7.png}%
	\caption{\scriptsize {Eigenvector 7 }}
\end{subfigure}
\begin{subfigure}{0.33\textwidth}
	\includegraphics[width=0.9\textwidth , height=0.56\textwidth]{Imgs-2/8.png}%
	\caption{\scriptsize {Eigenvector 8 }}
\end{subfigure}
\begin{subfigure}{0.33\textwidth}
	\includegraphics[width=0.9\textwidth, height=0.56\textwidth]{Imgs-2/9.png}%
	\caption{\scriptsize {Eigenvector 9}}
\end{subfigure}	
	\caption{\footnotesize Projection of image features along different eigenvector directions.}
	\label{fig:loss_pr}
\end{figure*}

\section{Reason for lower accuracy in CIFAR-10 and simulation results}
For CIFAR-10, even though the classification
accuracy is better than TripleNet [17], the performance is lower than the results reported by [10]. The reason for this might be that we use the grid search approach
for hyperparameter tuning. With Bayasian Optimization [12], we could find the
optimum hyperparameters of our model and this might improve the classification
accuracy of our approach in this dataset as well. For CIFAR-10, the batch sizes of $100$, $500$, and $1000$ and learning rates of $0.1$, $0.01$ and $0.001$ were used. The simulation results are summarized in Table~\ref{tab:simucifar-101}. The dropout values were $0.1$, $0.02$, $0.05$.
\par For STL-10, we measured the classification accuracy for batch sizes of $125$, $200$, and $250$ and learning rates of $0.001$, $0.002$, and $0.003$. We used dropout values of $0.1$, $0.3$, $0.5$ and $0.9$. The highest accuracy obtained on this dataset is $71.62\%$ which is the best value reported for LDA learning in STL-10 to the best of our knowledge. The results are summarized in Table~\ref{tab:simu2}.\par For MNIST, we used batchsizes of $100$, $500$, and $750$ and learning rate of $0.01$, $0.02$ and $0.03$. The dropout was varied from $0.1$, $0.3$, $0.5$, $0.9$. The simulation results are summarized in Table~\ref{tab:simumnist12}. The main point to be noted here is that since the number of classes is $10$, in all these datasets, our feature vector dimension is of length $9$ which corresponds to the LDA discriminant directions.

\begin{table}[t!]
	\begin{center}
		\begin{tabular}{|c|c|c|c|c|} % alignment of each column data
			\hline
			Lr & dropout & Accuracy in $\%$ (bs 100) & Accuracy in $\%$ (bs 500)  & Accuracy in $\%$ (bs 1000)\\ 
			\hline \hline
			0.1 & 0.1 & 81.22 & 75.73  & 65.72\\
			0.1 & 0.02 & 81.35  & 76.73  & 67.20\\
			0.1 & 0.05 & 82.23 &   78.23 & 68.21\\
			0.01 & 0.1 & 83.57  & 86.54  & 73.15\\
			0.01 & 0.02 &  83.98  & 87.32   & 73.11  \\
			0.01 & 0.05 &  84.23  & $\mathbf{88.17}$   &  72.17\\
			0.001 & 0.1 &  68.31  & 72.44  &  65.43\\
			0.001 & 0.02 & 70.22  & 71.98  & 66.97 \\
			0.001 & 0.05 & 72.20   & 65.21  &  68.23\\
			
			\hline     \end{tabular}
	\end{center}    
	\caption{\footnotesize Simulation results on CIFAR-10 for different batch sizes and dropout. Lr denotes the learning rate and bs denotes the batch size.}
	\label{tab:simucifar-101}
\end{table}

\begin{table}[t!]
	\begin{center}
		\begin{tabular}{|c|c|c|c|c|} % alignment of each column data
			\hline
			Lr & dropout & Accuracy in $\%$ (bs 125) & Accuracy in $\%$ (bs 200)  & Accuracy in $\%$ (bs 250)\\ 
			\hline \hline
			0.003 & 0.1 & 64.25 & $\mathbf{71.62}$  & 62.77\\
			0.003 & 0.3 & 58.17  & 67.41  & 57.69\\
			0.003 & 0.5 & 52.59 &   64.81 & 46.22\\
			0.003 & 0.9 & 41.19  & 63.71  & 42.77\\
			0.001 & 0.1 &  63.21  & 62.34   & 61.32  \\
			0.001 & 0.3 &  58.23  & 59.33   &  58.90\\
			0.001 & 0.5 &  55.22  & 57.68  &  55.98\\
			0.001 & 0.9 & 55.10  & 59.98  & 57.98 \\
			0.002 & 0.1 & 57.21   & 59.93  &  56.23\\
			0.002 & 0.3 & 55.32  & 58.58  & 54.43 \\
			0.002 & 0.5 & 53.29  & 56.64   & 54.23 \\
			0.002 & 0.9 & 52.59  & 57.71  & 55.55 \\
			\hline     \end{tabular}
	\end{center}    
	\caption{\footnotesize Simulation results on STL-10 for different batch sizes and dropout. Lr denotes the learning rate and bs denotes the batch size.}
	\label{tab:simu2}
\end{table}

\begin{table}[t!]
	\begin{center}
		\begin{tabular}{|c|c|c|c|c|} % alignment of each column data
			\hline
			Lr & dropout & Accuracy in $\%$ (bs 100) & Accuracy in $\%$ (bs 500)  & Accuracy in $\%$ (bs 750)\\ 
			\hline \hline
			0.01 & 0.1 & 98.71 & 99.57  & 99.43\\
			0.01 & 0.3 & 99.12  & \bf{99.73}  & 99.33\\
			0.01 & 0.5 & 99.61 &   99.53 & 99.55\\
			0.01 & 0.9 & 99.11  & 99.54  & 99.54\\
			0.02 & 0.1 &  98.73  & 98.45   & 98.23  \\
			0.02 & 0.3 &  97.34  & 97.98   &  97.55\\
			0.02 & 0.5 &  98.54  & 98.59  &  98.19\\
			0.02 & 0.9 & 99.06  & 98.47  & 98.89 \\
			0.03 & 0.1 & 98.43   & 98.33  &  98.56\\
			0.03 & 0.3 & 98.59  & 98.58  & 98.87 \\
			0.03 & 0.5 & 98.45  & 98.49   & 98.49 \\
			0.03 & 0.9 & 99.03  & 99.07  & 98.55 \\
			\hline     \end{tabular}
	\end{center}    
	\caption{\footnotesize Simulation results on MNIST for different batch sizes and dropout. Lr denotes the learning rate and bs denotes the batch size.}
	\label{tab:simumnist12}
\end{table}
	
	%% \paragraph{}
	
	\section{Implementation, Training and Performance Details}
	
.
	
	%  While the approach used for compositing the .
	
	% albedo, relight
	% wrinkles not rep. in mesh expressions
	% upsampled
	% realtime f2f, cpu
	% parallelized, optimized
	
	% \vfill\eject

	%% \paragraph{}
	
	
	%% \paragraph{}
	
	%% \input{implementation_details}
	%% \input{additional_results}
	%% \input{evaluations}
	% TODO: reference to the table. The caption should be fleshed out.